\documentclass[final,notitlepage,a4paper]{iopart}

\usepackage{iopams}
\usepackage{graphicx}
\usepackage{txfonts}

\begin{document}

\title[2D axi-symmetric Implicit electrostatic PIC/MC model I]{Implicit and electrostatic Particle-in-cell/Monte Carlo model in two-dimensional and axisymmetric geometry I: analysis of numerical
techniques}

\author{Hong-yu Wang$^{1,2}$, Wei Jiang$^1$ and You-nian Wang$^1$}

\address{$^1$School of Physics and Optoelectronic Technology, Dalian
University of Technology, Dalian, 116024, P.R.China}
\address{$^2$Depart of Physics, Anshan Normal University, Anshan,
 114007, P.R.China}
\ead{\mailto{ynwang@dlut.edu.cn}}

\begin{abstract}
We developed an implicit Particle-in-cell/Monte Carlo model in two-dimensional
and axisymmetric geometry for the simulations of the radio-frequency
discharges, by introducing several numerical schemes which include variable
weights, multigrid field solver, etc. Compared to the standard explicit models,
we found that the computational efficiency is significantly increased and the
accuracy is still kept. Numerical schemes are discussed and benchmark results
are shown. The code can be used to simulate practical reactors.
\end{abstract}

\submitto{\PSST} 
\vspace{2pc}
\maketitle
\section{Introduction}
Dc and rf discharges at low pressures, such as capacitively coupled plasmas
(CCP), inductively coupled plasmas (ICP) and magnetrons, have played critical
roles as etching and depositing devices in semiconductor industry
\cite{Lieberman05,Makabe06}, as well as in some other applications, such as
plasma lighting, displays, healthcare and Hall thrusters \cite{Kushner08}.
Computer simulations have been demonstrated as a powerful tool in this field.
They can give unique insights into the fundamental plasma physics and reduce
the workload for the industrial reactor designers significantly.

There are three commonly used simulation techniques, namely, the fluid,
Particle-in-cell(PIC), and Boltzmann model in plasma physics research
\cite{Kim05, Iza07, Dijk09}. PIC model \cite{Hockney81, Birdsall85} solves the
Newton and Maxwell equations directly. Kinetic, non-local and non-equilibrium
effects can be included. In addition, Direct Simulation Monte Carlo (DSMC)
method \cite{Bird94} is used for modeling rarefied neutral gas flows, in which
the mean free path of a molecule is on the order of (or greater than) the
characteristic physical scale. Both PIC and DSMC models need to be coupled
together to depict these discharges. This method, often referred as PIC/MC
model, was firstly introduced at early 1990¡¯s
\cite{Vender90,Surendra90,Birdsall91}. PIC/MC simulations for these discharges
adopted simplified model from pure PIC model and DSMC method. In many
discharging systems, electrostatic or Darwin modeling are sufficient. At the
same time, because the plasma density and gas pressure is low, the neutral
molecules are in their own thermal equilibrium, and the Coulomb collision rate
is relatively low. Therefore one needs only to consider the collisions between
the charged particles and the neutral molecules. Null collision method had been
proved to be an effective method to treat these MC processes, in which one need
not to scan all the particles, as it often does in DSMC simulations.

The main difficulties of PIC/MC simulations are the costs of computational
resources. In conventional electrostatic PIC simulations, the spatial and
temporal steps must be chosen to resolve the fastest temporal and finest
spatial behavior of the electrons, namely, $\Delta x \leq \lambda_{D}$, $\omega
_{p} \Delta t \leq 2$ and ${\Delta x}/{\Delta t}<v_t$, where $v_{t}$ is the
electron thermal velocity. This would require hundreds to thousands cells per
cm and the time step of $10^{-12}\sim 10^{-11}$s in these discharge modeling.
On the other hand, typically more than 100 particles per cell are needed to get
rid of the stochastic errors in MC process \cite{Turner06}. Since the
computational costs are very high, most PIC/MC simulations were only done in 1D
geometry up to now. Conventional 2D or 3D simulations are only possible for
some cases where the densities are relatively
low\cite{Yonemura06,Wu07,Lisovskiy10}. For most higher density cases in CCP,
PIC/MC simulations of practical interesting systems often run on supercomputers
\cite{Wakayama03,Wu07,Wang09}. To overcome this problem, some fast but
non-standard algorithms were proposed, such as the global PIC/MC method
\cite{Yan99} and fluid PIC \cite{Lapenta95,Soria09}. However, they made
additional approximations and may not be sufficient for investigating of some
kinetics effects, for example, when the distribution functions of electrons are
anisotropic \cite{Jiang09}.

Fortunately, the fastest phenomena are usually not very important in these
discharges. As the fastest oscillation modes of electrons are not very
important here, we only need solve rf frequency $\omega_{rf}$ plus some
harmonics. By damping some high frequency modes, implicit PIC/MC method can
eliminate the major constrains of grid spacing and time step on explicit PIC
codes, while most of the kinetics effects are still kept, and thus it could be
a better approach to these problems. However, for implicit model, more
complicated algorithms must be introduced, and the numerical schemes should be
carefully treated, especially when coupling with MC model and in axisymmetric
geometry. Here are the main difficulties origin from the cylindrical geometry
listed below: (1) weighted particles resizing; (2) implicit particle pushing in
cylindrical geometry; and (3) Poisson solver and its parallelization in
cylindrical geometry, especially when $R \gg L$, where R is the device radius
and the L is the electrode spacing. There are many solutions to them for the
explicit simulations, For example, Nanbu
\cite{Yonemura06,Wakayama03,Denpoh98,Nanbu99,Nanbu00,Nanbu01,Takekida04,Takekida05}
has solved nearly all essential problems associated with the MC method and the
axisymmetric geometry for explicit algorithms by developing some new methods
and introducing some methods from DSMC model. But for the implicit simulations
with MC model, subtle difficulties exist still.

This is the first of our two serial papers. In this paper, we report a direct
implicit and electrostatic PIC/MC simulation model for CCP in two-dimensional
axisymmetric geometry. Although our model is designed for CCP, it can be used
to study some other problems, such as dc and atmosphere discharges. This work
would not have been possible without those who have developed the PIC and DSMC
method to their present advanced state. We should emphasize that most of the
numerical techniques here have been developed by many previous researchers,
many of which have histories of more than ten years. We just try to incorporate
these algorithms together, then analyze and compare all possible numerical
treatments to meet the specific requirements. The PIC algorithms mainly follow
up the works by Birdsall \cite{Birdsall85,Birdsall91}, Langdon and Cohen
\cite{Langdon82,Langdon83,Brackbill85}, Verboncoeur \cite{Verboncoeur93,
Verboncoeur01, Verboncoeur05}, Vahedi \cite{Vahedi93a,Vahedi93b,Vahedi97},
Hewett \cite{Hewett87,Hewett92} and Friedman \cite{Friedman90}. The MC
algorithms mainly follow up the works by Bird \cite{Bird94}, Nanbu
\cite{Nanbu00} and Vahedi \cite{Vahedi95}. We will discuss these numerical
schemes in Sec.2. Simulation results and benchmarks are given in Sec.3.
Discussions and a brief summary are presented in Sec.4.

\section{Numerical Methods}

\subsection{Direct Implicit PIC Simulation}
In the Implicit Particle-In-Cell (IPIC) schemes, the field in the next step
$E^{n+1}$ which depends on the future charge density $\rho^{n+1}$ must be known
at $n$ step. There are two kinds of algorithms, namely, direct implicit
simulation (DIPIC)\cite{Langdon82,Langdon83,Hewett87,Friedman90} and implicit
movement method simulation (IMMPIC) \cite{Mason81, Brackbill82, Vu92,
Lapenta06}. In DIPIC, the field equations are derived from direct summation and
extrapolation of the particles moving equations. In the implicit movement
method, the field equations are derived from the Vlasov movement equations.

Here we applied the direct implicit simulation method which was proposed by
Langdon\cite{Langdon82} and Friedman \cite{Friedman90}. In brief, in "D1"
electrostatic DIPIC algorithm, the particle pushing procedure is divided into
"firstpush":

\begin{eqnarray}
\nonumber\tilde{v}^{n+1/2}&=&v^{n-1/2}+\frac 12 \bar{a}^{n-1}\delta t\\
\nonumber\tilde{x}^{n+1}&=&x^{n}+\tilde{v}^{n+1/2}\delta t
\end{eqnarray}

 and "finalpush":
 \begin{eqnarray}
\nonumber x^{n+1}&=&\tilde{x}^{n+1}+\frac 12 a^{n+1}\delta
t^2\\
\nonumber v^{n+1/2}&=&\tilde{v}^{n+1/2}+\frac 12 a^{n+1}\delta t
\end{eqnarray}
where
\begin{equation} \label{accu}\bar{a}^n=\frac 12 (\bar{a}^{n-1}+a^{n+1})
\end{equation}

Between the two pushing procedures, electric field in time $t^{n+1}$ is solved by
\begin{equation}
\nabla\cdot[1+\chi(\vec{x})]\nabla \phi^{n+1}=-\tilde{\rho}^{n+1}
\end{equation}

where $\tilde{\rho}^{n+1}$ denotes the charge density contributed by
$\tilde{x}^{n+1}$, and
$$\chi(\vec{x})=\sum_{\nu}\frac 12
\tilde{\rho}^{n+1}_{\nu}\frac {q_\nu}{m_\nu}\delta t^2$$

The $\sum_{\nu}$ denotes summation over all species particles.

After the pushing and field solving procedures are executed, the MC procedure is
executed.

In summary, the simulation cycle consists of the following steps: (1) first
pushing; (2) weighting; (3) solve the field equation; (4) final pushing; and
(5) MC process. The first pushing in $(n+1)_{th}$ cycle and final pushing in
$n_{th}$ cycle can be merged into one procedure, so only one passing through
the particles is needed, which will improve the code efficiency. The only
difference is the MC procedure parameters ($x^{n}$ or $\tilde{x}^{n}$). Our 1D
and 2D benchmarks show that both methods produce identical results.

\subsection{Unweighted and weighted particles}
Unlike the Descartes coordinates, even the grid spacings and the densities are
uniform, the volumes of the grid cells are proportional to the radius in
cylindrical coordinates. In this case, even if super particles were assigned to
identical number of physical particles and the grids were unform with identical
macro particle numbers per cell, they would not give constant density.

In general, we can apply two methods, i.e., unweighted particles and weighted
particles. One can adopt non-uniform grids along R for unweighted particles
\cite{Nanbu99}, but this algorithm should lead to very large grid spacing near
the axis which will produce large errors in implicit schemes and Poisson
solver. One can also use uniform grid spacing and set the particles number in
one cell proportional to the radial position. However, MC process requires that
the super particles number in one cell should not be too small. This method
will either produce small number of the particles near the axis to disturb the
MC process or lead to excessive large particle numbers in the outer grid cells.
So unweighted particles are not recommended here.

To overcome this problem, one can adopt weighted particles \cite{Bird94,
Takekida04,Lapenta02,Chanrion07,Hoyo08}. A certain weight $w_p$ is assigned to
each particle according to its radial position or the volumes of the cell:
\begin{equation}
w_p\propto 2\pi r \Delta r \Delta z
\end{equation}
Here $w_p$ is the number of the physical particles which is contained by one
super particle. This method will make identical numbers of the super particles
in each cell when the initial densities and the grids are uniform. The major
problem associated with this scheme is that the super particle needs to be
properly resized when it moves radially.

In DSMC simulations, Bird \cite{Bird94} presented zero-order resizing method,
where the super particle numbers were changed according to their positions but
momentums and energies were kept. The new weighting factors may be based on
either the radius of the cell or the the radius of the particle itself. In the
particle based weighting scheme, when a super particle moves from radial
position $r$ to $r'$ in one step, it has the possibility to be discarded or
duplicated according to a certain probability to ensure average charge
conservation. Cell based weighting scheme is similar, except that the particle
is discarded or duplicated when it crosses the cell boundaries. Cell based
weighting scheme is introduced by Nanbu \cite{Takekida04} from DSMC model, and
is successfully used for inductive coupled plasma simulations
\cite{Takekida05}. However, cell based weighting is not recommended
\cite{Bird94}, because it will lead to errors when particles moves parallel to
the axis, and it can not maintain the smooth flow gradients normal to the axis.

In these resizing processes the charges only conserve on average, which would
produce the random walks problem. This problem has been well known for long in
DSMC simulations \cite{Bird94}. Random walks is not a major problem in DSMC
simulations, because the molecular interactions are short range force and there
are no interactions in the free flight phase. In PIC/MC simulations, the
dominating electric force is a long range force and the collective electron
oscillation in radial direction exists. Then the fluctuation of electron
density would bring electrostatic waves and nonphysical effects.

It would be more accurate that the super particle number is conserved, i.e.,
particle weight will not be changed when it moves. Some PIC codes
\cite{Lapenta02, Chanrion07,Hoyo08} adopted this scheme. DSMC method also
adopted similar treatment, and one often referred to this scheme as stochastic
weighted method \cite{Rjasanow96}. However, the small weight particle may be
replaced by the larger weight particle. After the simulation runs some rf
periods, we find that the super particle numbers near the axis tend to
proportional to the radius. There would be only 10's particles with large
weights near the axis after running some periods. This effect is more obvious
for electrons, which will bring nonphysical heating in the axis. After the
simulation runs some periods, the average weights of the particles become
larger. So we adopted a special particle split scheme. After the code runs some
periods, the weights of the particles $w_p$ are checked and compared with the
weights $w_r$ calculated from their present radial position. If $w_p>w_r$, the
particles are split to several new particles ($N=[w_p/w_r]$). The positions,
velocities and accelerations of the new particles are duplicated from the old
ones. The weights of the new and old particles are set to $w_r$ and $w_p-Nw_r$
respectively. Here all phase space information is kept and the charge is
conserved. After the system reaches equilibrium, this splitting method will
only change the numbers of the super particles, but not change the plasma
density and the field. This method will increase the super particle numbers, so
sometimes one need to combine the small particles \cite{Lapenta02}.

We have benched all three weighting assigning schemes. When a small number of
particles is used, especially at small radius, we find that particle based
resizing method tend to produce larger density than weighting-conserving
scheme.  With enough large number of particle per cell, both methods produced
similar results. But for cell based method, the radial density is not very
smooth and many small peaks appear in the density profile, which implies that
additional electrostatic modes are excited and thus larger density can be
generated. Therefore we recommend the weighting-conserving scheme with enough
particles per cell. When using zero-order resizing scheme, we do not use the
global buffer like the DSMC method \cite{Bird94, Liechty08}, but just duplicate
the particles.

\subsection{Particle Pushing}
   In the first and final pushing steps, the accelerations, velocities and positions of
the particles are updated. In Descartes coordinate, it is straightforward:
every position component should be added with the velocity component multiply
$\Delta t$ while every velocity components should be updated accordingly. But
in curvilinear coordinate systems, the position components could couple
together. It is not recommended to push the particles by using the moving
equations in cylindrical coordinate: using $\Delta \theta=v_{\theta}\Delta t/r$
will produce larger $\Delta \theta$ at small $r$, then one needs adopt smaller
$\Delta r$ and $\Delta t$ at small $r$. A feasible way is local coordinate
transformation. Both DSMC method and PIC code \cite{Birdsall85,Bird94,
Wallace86} have adopted this treatment, but here we need some modifications.
The particle are described by the parameters
$\{r,z,v_r,v_\theta,v_z,a_r,a_z\}$. Then the first pushing can be
applied as follows :
\begin{eqnarray}
\nonumber v_x'=v_r+\frac 12 a_r\Delta t\\
\nonumber v_y'=v_\theta+\frac 12 a_{\theta}\Delta t\\
\nonumber x'=x+ v_x'\Delta t \\
\nonumber y'=v_y'\Delta t\\
\nonumber v_z'=v_z+\frac 12 a_z\Delta t\\
z'=z+v_z'\Delta t
\end{eqnarray}
Then the coordinates are rotated, the new velocities and
accelerations are given by
\begin{eqnarray}
\nonumber
\{v_r',v_\theta'\}=\{v_x'\cos\theta+v_y'\sin\theta,-v_x'\sin\theta+v_y'\cos\theta\}\\
\nonumber \{ a_r', a_{\theta}'\}=\{a_r\cos\theta,-a_{r}\sin\theta\} \\
a_z'=a_z
\end{eqnarray}
The final pushing is executed in full X-Y coordinates($i=r,z$):
\begin{eqnarray}
\nonumber x_i''=x_i'+\frac 1 2 \frac q m E_i \Delta t^2\\
\nonumber v_i''=v_i'+\frac 1 2 \frac q m E_i \Delta t\\
\nonumber a_i''=\frac 12 (\frac q m E_i+a_i')
\end{eqnarray}

When particles hit on the electrodes,we remove them from the moving particle
lists and add the charges to the depositing charges of the electrodes. If
particles pass the axis, the position, velocity and acceleration of the
particles in R direction are changed to their absolute values. In this scheme
we have ignored the $a_\theta$ damping accumulating (equ [{\ref{accu}}]). If
time steps are large, there could be some errors on $v_\theta$ at small radius.
However, the differences in our cases are neglectable.

  The other natural way is to adopt global Descartes (X-Y-Z) coordinate, in
which every particle has its Descartes coordinates, velocities and
accelerations in full three dimensions despite that the field is still in two
dimensions. The first pushing can be done in Descartes coordinate. In the final
pushing, we weight the $E_r$ with the particle position radius
$r=\sqrt{x^2+y^2}$ and calculate the Descartes components of electric field,
\begin{equation}
\label{xyz}
\tan \theta=\frac {x}{y}\\
 E_x=E_r \cos\theta\\
 E_y=E_r \sin\theta
\end{equation}
Then the final pushing is executed in full X-Y-Z coordinates. If r is very
close to zero, we set $\cos{\theta}=1$ and $\sin{\theta}=0$ in Equation.
\ref{xyz}.

In the particle initialization, we get the particles radius $r_p$
uniformly ($r_p=p/N *R$, $N$ is the total particles number) then
$x_p$ and $y_p$ are given by
\begin{equation}
x_p=r_p \cos{\theta}, \\
y_p=r_p \sin{\theta},
\end{equation}
where $\theta$ is randomly sampled between 0 and $2\pi$.

We benchmarked above two algorithms and find no obvious differences in the
results. The local X-Y scheme run slightly faster than the global X-Y-Z scheme.

\subsection{Weighting}

In curvilinear coordinate systems, assigning the particle charge to the grid
must be specially treated, which has been well studied by Verboncoeur
\cite{Verboncoeur01,Verboncoeur05} in the most generalized form. There are two
frequently-used weighting methods in cylindrical coordinates : bilinear
weighting in $z-r$ and bilinear weighting in $z-r^2$.  In $z-r$ weighting, the
real particle number $N_{i,j}$ assigned to the grid point at $(r_i,z_j)$ can be
written as
\begin{equation}
N_{i,j}=w_p\frac{(z_{i+1}-z_{p})(r_{j+1}-r_{p})}{(z_{i+1}-z_{i})(r_{j+1}-r_{j})}.
\end{equation}

The particle number $N_{i,j}$ assignment in $z-r^2$ weighting is
\begin{equation}
N_{i,j}=w_p\frac{(z_{i+1}-z_{p})(r_{j+1}^2-r_{p}^2)}{(z_{i+1}-z_{i})(r_{j+1}^2-r_{j}^2)}.
\end{equation}
Here $r_p=\sqrt{x_p^2+y_p^2}$ for X-Y-Z coordinates. The density
$n_{i,j}$ is calculated by
\begin{equation}
n_{i,j}=\frac {N_{i,j}}{V_j}
\end{equation}
where $V_j=\frac 13 \pi
[r_{j+1}(r_j+r_{j+1})-r_{j-1}(r_j+r_{j-1})](z_{i+1}-z_{i})$. and $V_0=\frac 13
\pi r_1^2 (z_{i+1}-z_{i})$ for $z-r$ weighting; $V_j=\frac 12 {\pi}
(r_{j+1}^2-r_{j-1}^2)(z_{i+1}-z_{i})$ and  $V_0=\frac 12 \pi r_1^2
(z_{i+1}-z_{i})$ for $z-r^2$ weighting. The weighting of field $E_r$ and $E_z$
acted on the particle are done similarly. Our benchmarks show that those
different weighting schemes give very similar results.

\subsection{The field solver on the cylindrical coordinate systems}
In electrostatic DIPIC, one of the key issues is to construct a fast and stable
Poisson solver. The solver should have good scalability when being
parallelized. In this problem, it must be suitable for the case of $R\gg Z$ to
deal with the real-size CCP reactors.

There were many Poisson solvers that can deal with the variant coefficient
Poisson equation. Because of the variable dielectric constant
$\epsilon=1+\chi$, the fast Poisson solver (based on the Fourier Transform or
Cyclic Reduction \cite{Buzbee70}) can not be applied directly. Historically,
some researchers have used global iterations to construct the solver
\cite{Brackbill85}. These solvers can be constructed only by the Fast Poisson
Solver, but they have variable convergence rates. When the dielectric
coefficient $\chi$ varies, the iterating times of the solver can increase to
unacceptable level. Goloub et. al \cite{Golub73} transformed the Poisson
equation to a Helmholtz equation and solved it by a similar iteration solver.
However, the solver has similar shortcoming. The Dynamic Alternating Direction
Implicit (DADI) algorithms \cite {Hewett92} can be applied here but showed low
efficiency. Typical Krylov iteration methods (Conjugate Gradient or GMRES, etc
) have similar shortcomings. Recently, some authors use Krylov procedures to
deal with the Helmholtz equation from Goloub's method \cite{Nelson01}. The
major shortcomings of these algorithms are the complexities.

In our case, because of the positivity of the $\chi$, the equation is a
negative elliptical equation. The equation can be discretized by the finite
volume scheme\cite{Birdsall85}. The only difference is the susceptibility at
half integer point be selected to $\chi_{i+1/2,j}=\frac 12
(\chi_{ij}+\chi_{i+1,j})$ and $\chi_{j,j+1/2}=\frac 12
(\chi_{ij}+\chi_{i,j+1})$, or $\chi_{i+1/2,j}=\max(\chi_{i,j},\chi_{i+1,j})$
and $\chi_{i,j+1/2}=\max(\chi_{i,j},\chi_{i,j+1})$. Our 1D benchmarks didn't
show any differences between the two methods.

Consider the simple finite volume discrete scheme\cite{Birdsall85} or similar
five-point discrete scheme, the discrete Poisson equation has the form
\begin{eqnarray}
\nonumber -a_{i,j}\phi_{i-1,j}+b_{i,j}\phi_{i,j}-c_{i+1,j}\phi_{i,j}\\
-d_{i,j}\phi_{i,j-1}-e_{i,j}\phi_{i,j+1}=h^2
f_{i,j}
\end{eqnarray}

In uniform grids, this scheme has two order accuracy. All of the coefficients
(a,b,c,d,e) are positive and we have $b_{ij}=a_{i,j}+c_{i,j}+d_{i,j}+e_{i,j}$
for the uniform grids. In addition, because the $\chi$ depends on the charge
densities, the $\chi(x,y)$ distribution is smooth spatially. So the multigrid
method \cite {Hackbush85,Trottenberg01} is a good choice to construct the
solver. However, typical etching devices have cylindrical shapes and the radius
are much larger than the height. In cylindrical systems, the standard multigrid
solver can cause slow convergence rates: The convergence rates of typical
Descartes multigrid 2D Poisson solvers are constants and less than 0.1, which
means about $8-10$ V cycles reduce the error norm to $10^{-10}$. For $64*64$
$r-z$ cylindrical Poisson systems, the iterations increase to about 30
\cite{Iyenger90}. When the grid numbers on r direction increase, the converging
speed becomes much lower. Additionally, because the typical discharge device is
large, we need parallelize the code. The parallelization of standard multigrid
solvers is complex and case dependent. So we developed a semi-coarsening
multigrid solver \cite{Trottenberg01}.

\begin{figure}
\includegraphics[scale=0.32,angle=-90,bb=0 0 500 400]{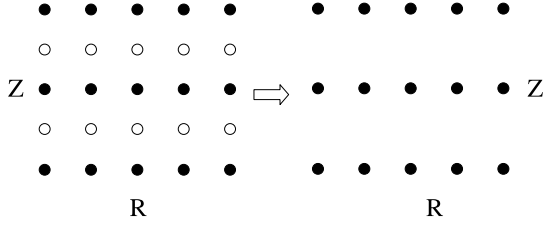}
\center \caption{The concepts of the grid z-coarsening}\label{mg}
\end{figure}

Semi-coarsening multigrid procedure has been applied to the anisotropic problem
\cite{Schaffer98,Lai07}. If the grid spacing on one direction is much less than
the other directions, one can coarsen this direction grid only. On the other
hand, the grid in our problems is uniform but we run the z-coarsening only. The
concepts of z-coarsening are showed in Figure {\ref{mg}}. After one turn
coarsening, the grid spacing in z direction is doubled.

When z-coarsening is applied, the coarsened differential stencil will become
very anisotropic, so standard Gauss-Seidel smoothing is not effective. To
overcome the difficulty, we apply a line-smoothing procedure: write the
coarsened equations as
$$-a_{i,j}\phi_{i-1,j}+b_{i,j}\phi_{i,j}-c_{i,j}\phi_{i+1,j}-d_{i,j}\phi_{i,j-1}-e_{i,j}\phi_{i,j+1}
=h^2 f_{i,j}$$
where $i=1,M$ along the z-direction and
$j=1,N$ along the r-direction. The line smoothing is executed by
$i=1,M$. For every i, the equation sets are written to
\begin{eqnarray}
\nonumber &&......\\
\nonumber
&-&d_{i,j-1}\phi_{i,j-2}^{new}+b_{i,j-1}\phi_{i,j-1}^{new}-e_{i,j-1}\phi_{i,j}^{new}=\\
\nonumber&&=h^2
f_{i,j-1}+a_{i,j-1}\phi_{i-1,j-1}^{old}+c_{i,j-1}\phi_{i+1,j-1}^{old}\\
\nonumber
&-&d_{i,j}\phi_{i,j-1}^{new}+b_{i,j}\phi_{i,j}^{new}-e_{i,j}\phi_{i,j+1}^{new}=\\
\nonumber&&=h^2
f_{i,j}+a_{i,j}\phi_{i-1,j}^{old}+c_{i,j}\phi_{i+1,j}^{old}\\
\nonumber
&-&d_{i,j+1}\phi_{i,j}^{new}+b_{i,j+1}\phi_{i,j+1}^{new}-e_{i,j+1}\phi_{i,j+2}^{new}=\\
&&=h^2
f_{i,j}+a_{i,j+1}\phi_{i-1,j+1}^{old}+c_{i,j+1}\phi_{i+1,j+1}^{old}
\end{eqnarray}
When the equation set is solved, the $i_{th}$ line smoothing has been executed.
In every smoothing procedure, the line smoothing is executed from $i=1$ to M
with Red-Black order and the grid values of $\phi$ are updated.

Numerical tests show that V(0,1) cycles provide good convergence rates. The
algorithm can be described like follow:

Algorithm 1. Multigrid V-cycle MG(u,f)
\begin{eqnarray}
\nonumber&u^{L}&=w_0 \\
\nonumber do &until& converge\\
\nonumber&f^{L}&=f\\
\nonumber&do& \ l=L-1,2\\
\nonumber&&u^{l} = 0\\
\nonumber&&f^{l-1} = I^{l-1}_l (f^l-A^l u^l)\\
\nonumber&&A^{l-1}=I^{l-1}_l A^l\\
\nonumber&enddo&\\
\nonumber&solve& A^1 u^1=f^1\\
\nonumber&do& \ l=2,L \\
\nonumber&&u^{l} =u^l+ I^{l}_{l-1} (u^l)\\
\nonumber&&linesmooth(u^l,f^l)\\ \nonumber&enddo&\\
 \nonumber loop &&
\end{eqnarray}

The restrict and prolong operators are simple line forms:

\begin{equation}
I_{l-1}^l=[1/4 \ 1/2 \ 1/4]^T
\end{equation}

We benchmark the solver and find that when $\chi=0$ the solver
converges constantly for varied size grids with Dirichlet boundary
condition. When $\chi \sim 1$ with smoothing spatial distribution,
the solver converges to $10^{-10}$ by $L^\infty$ norm after about 12
times V-cycles.

Typically, the modeling should consider the external circuit of the reactors.
For example, dc self-biasing voltage can be built up on the electrically
powered electrodes due to the blocking capacitor. The general way to include
these effects is given by Verboncoeur \cite{Verboncoeur93}. The surface charges
are included in the Poisson equation, coupled with the external circuit. In our
cases, since the coupling between the electrodes is not strong, we have adopted
the Vahedi's model \cite{Vahedi97}, which is the simplified version of
Verboncoeur' method and can be numerically realized easily. In this model, the
electric fields with external circuit can be obtained by linear superposition
of the solutions to two problems: At first, the Poisson equation with exact
charge density and a zero boundary condition is solved:
$$\nabla\cdot(\chi\nabla \phi_P)=-\rho$$
$$\phi_P|_{electrode}=0$$

Secondly, a Laplace equation with same coefficient and normalized boundary
condition is solved. For example, if the upper electrode is grounded and RF
power is applied on the lower electrode, the Laplace problem will be
$$\nabla\cdot(\chi\nabla \phi_L)=0$$
$$\phi_L|_{upper}=0,\ \phi_L|_{lower}=1$$

Finally, the voltage $V^t$ of the powered electrodes are calculated by solving
the circuit equation and get the field solution. For example, when a capacitor
is applied between the RF source and the powered electrode, the circuit
equation becomes:
\begin{equation}
Q^t=Q^{t-1}+C[V_{rf}^t-V^t]-Q_c^{t-1}+Q_{conv}^t,
\end{equation}
where $Q_{conv}^t$ is the charge deposited on the RF electrode from $t-1$ to t,
$Q$ and $Q_c$ are the charges deposited on the electrode and the capacitor,
while $V_{rf}^t$ is the instantaneous rf source voltage. However, because the
solver is run with the $\tilde{x},\tilde{v}$, the $Q_{conv}$ in the real
solvers is calculated from this step's first pushing and the previous step's
final pushing. Because the self-biasing is a slow varying process, this causes
no observable errors. Then we get the $\phi$ by $\phi=\phi_P+V^t \phi_L$. We
need to solve more Laplace equations for the extra RF-applied electrodes.

One can also include the external circuit effects by physics insights
\cite{Yonemura06,Wakayama03}. This method is to force the net current flowing
into the electrode to zero in one rf period. One can adjust $V_{dc}$ within the
voltage waveform of $V=V_{rf}\sin\omega t+V_{dc}$ to satisfying this condition.
This method is very easy to be incorporated, but it is not recommended here
because of its narrow applicability.

\subsection{Monte Carlo model}

Since the Monte Carlo model is a well-developed method, we only briefly discuss
the numerical treatments. At present, our codes only include gas models for Ar,
O$_2$ and CF$_4$. For electrons, we adopted the null collision method and the
molecular velocity is assumed to be zero since $v_e \gg v_n$. For the
non-reactive collisions between ion and neutral gas, null collision method is
still adopted, and the molecular velocity is sampled from Maxwellian
distribution. The velocities of the electrons and the ions after non-reactive
collisions are given by Vahedi' method \cite{Vahedi95}. The velocity of the ion
after a reactive collision is given by Nanbu and Denpoh's method
\cite{Denpoh98,Nanbu00}. The Ar cross sections come from \cite{Phelps99} while
the $\tt{O_2}$ cross sections come from \cite{Vahedi95}, and the $\tt{CF_4}$
cross sections come from the BOLSIG package \cite{Boeuf}. All of them are
linearly interpolated. For the energy being higher than the available data, we
extrapolated the cross sections by the $1/E$ law \cite{Zecca01}.

The standard sampling procedure of the null collision method is still adopted
here, regardless the different weights of the particles. We have also proposed
weighting-based sampling procedure for the null collision method. Our 1D
benchmarks of both methods have showed that, weighting-based sampling procedure
is more accurate, especially when larger density gradient exists. However,
there is only up to $10\%$ difference between the two sampling procedures, and
the weighting-based sampling procedure runs slower and has some problems in 2D
problems. So we still use the standard sampling procedure in our present
research, and we will discuss the new method elsewhere.

In the Monte Carlo processes in which the new particles are generated, the
weights of the new particle is just duplicated from the incident particles. The
accelerations $\bar{a}$ of the new particles must be set to a reasonable value.
We set the accelerations according to the charge mass ratio:
\begin{equation}
\bar{a}_{i}=\bar{a}_{j} \frac {q_{i}m_{j}}{q_{j}m_{i}}.
\end{equation}.

\subsection{Speeding up and parallelization}

There are many ways to speed up the code\cite{Wang09,Kawamura00, Tskhakaya07}.
We have adopted subcycling in our code with Vahedi's method\cite{Vahedi93a},
and speed boost nearly 2 is achieved for 1D cases. Particle sorting, which is
very successful in our explicit code \cite{Wang09}, shows no speeding up and
sometimes it is even more slow than the unsorted cases. The reason is partially
because the particle sorting is a time consuming operation, and partially
because the grid sizes are much smaller than the explicit simulation and thus
the cache missing is not very serious.

On the other hand, although implicit algorithm can be executed by much smaller
grid numbers than explicit methods, the simulation particles number is still
very large. So we need parallelize the code \cite{Wang09}. Here we adopt the
MPI\_ALLReduce framework\cite{Kawamura00, Wang09}: $\chi$, $\rho$ and
$Q_{conv}$ are summed up to each processor, the serial Poisson solver are
executed on every processor to calculate the fields. Since the simulation size
is not very large here, the communication time and Poisson equation solving
time are very little. So the parallel efficiencies are satisfying.

\section{Benchmarks and results}

All simulations are carried out on our 12 nodes PC clusters: each node has an
Intel Core2 E4500 CPU and 2G memory. The nodes are connected by 1000M ethernet
networks. The clusters have about 210G FLOPS R$_{peak}$ and about 110G FLOPS
linpack R$_{max}$. We normally run two processes in one CPU for 2D parallel
simulations.

The physical parameters of the benchmark problems are listed as follows: The
frequency of rf source $\omega_{rf}=2\pi 13.56$MHz. Voltage source is applied
to the electrode at $z=0$cm with the waveform of $V_{rf}=200 \sin
\omega_{rf}t$. Argon gas is used with the pressure of 100mTorr and the
temperature of 300K. The electrodes spacing is $2cm$ while the radius is $8cm$
and the gap between the lower power electrode to the grounded outer cylinder is
$2cm$. Here we do not consider the self-biasing effect. The initial density is
uniform of $5\times10^{15}$m$^{-3}$ for all cells and 200 particles are placed
randomly within one cell. All simulations are run for 1000 rf periods, and all
results below are given by averaging one rf period.

\subsection{1D results}

The 1D simulations are performed to benchmark the implicit results with the
explicit simulations and to determine the space and time steps. The 1D
simulations are run serially in one node and will only take about several ten
minutes for implicit code (with 64 cells and  $\Delta t=0.5\times10^{-10}s$ )
and about several ten hours for explicit code (512 cells and $\Delta
t=1.25\times10^{-11}s$).

\begin{figure}
\includegraphics[scale=0.5]{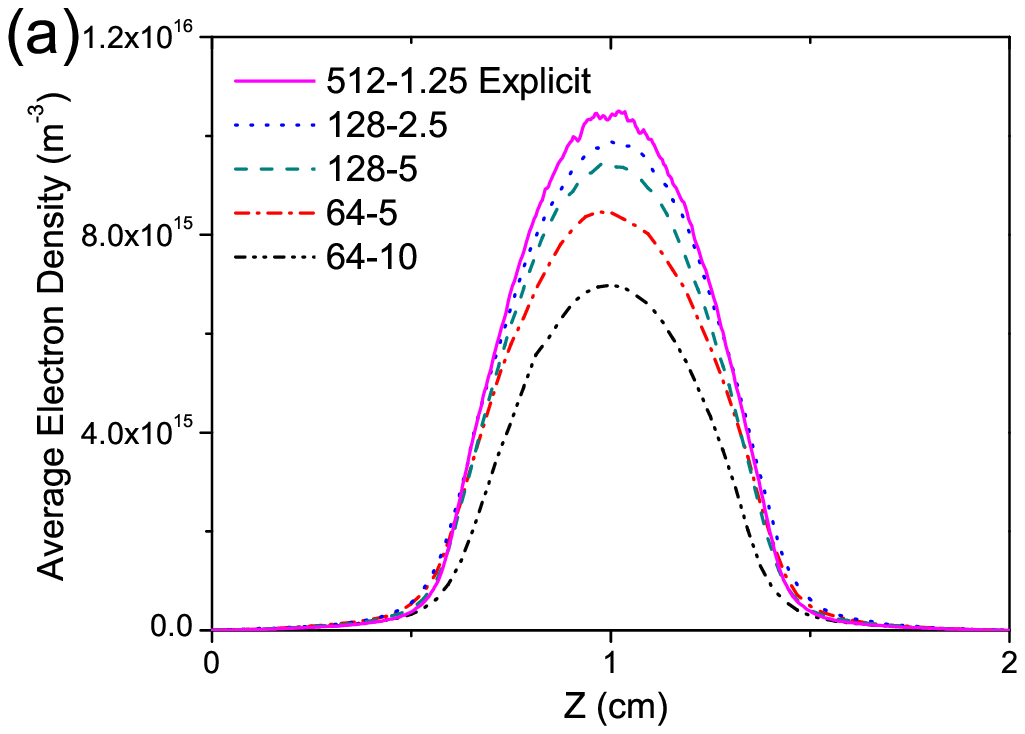}
\includegraphics[scale=0.5]{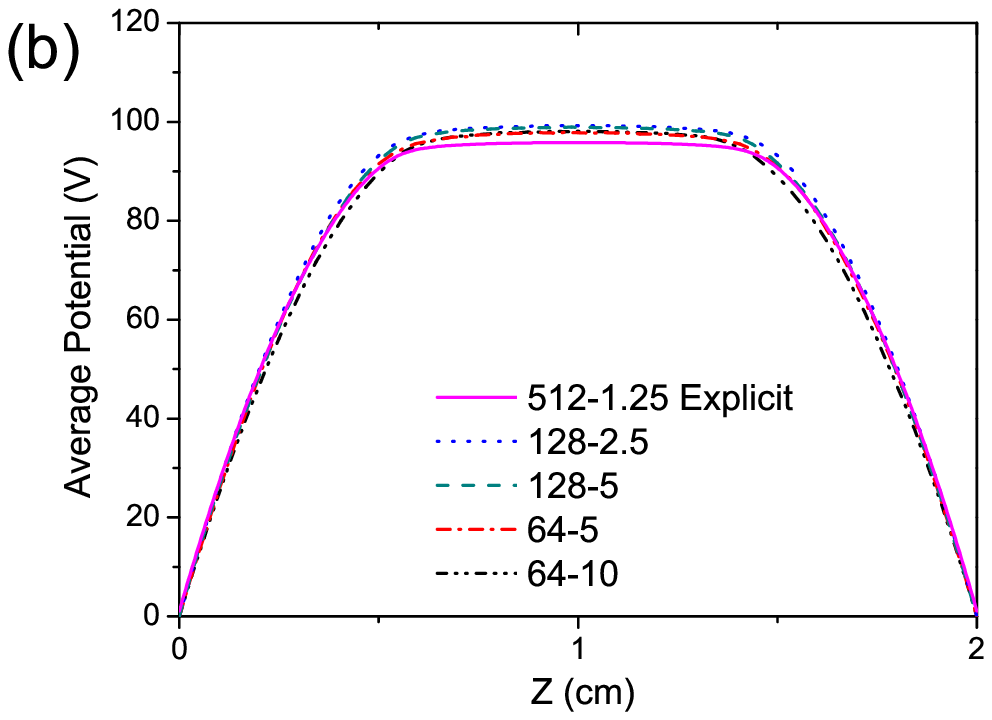}
\caption{ Simulation results with different space and time steps
(the former number is the number of the grid and the later one is the time steps
$\times10^{-11}s$): (a) average electron density profiles ; (b)
time averaging potential.} \label{1dres}
\end{figure}

Firstly we compare the results of implicit and explicit algorithms. Fig.
\ref{1dres}a shows the average electron density profiles with different spacing
and time steps. For the implicit numerical schemes, we have adopted
$\chi_{i+1/2,j}=\frac 12 (\chi_{ij}+\chi_{i+1,j})$ and $\chi_{j,j+1/2}=\frac 12
(\chi_{ij}+\chi_{i,j+1})$. From Fig. \ref{1dres}a, we can clearly see that
implicit schemes can give reasonable densities. However, the densities from the
implicit schemes are lower than those from the explicit code. Larger space and
time steps tend to give smaller center density and smaller bulk plasma length
due to the excessive damping of the high frequency modes. We also find that the
damping error is more sensitive to the time steps than the space steps. This is
beneficial to the simulations because the computational cost is inversely
proportional to the square of the grid size and is only proportional to time
steps. This would allow us to use larger space step and smaller time steps
while keeping the accuracy. It seems that $N_z=64$ and $\Delta
t=0.5\times10^{-10}s$ is sufficient to the simulations, because the bulk plasma
length is nearly identical to that of the explicit one and the plasma density
is only about $20\%$ lower. We considered this difference is acceptable,
because most physics involved here is still kept. One can also adopt finer
space and time spacing to reach better accuracy but cost more running time.

Fig. \ref{1dres}b shows time averaging potentials from the same parameters.
There are little differences between different space and time steps, and
potential of the explicit scheme is slightly lower than that of the implicit
one. It can be seen that over a wide range, the implicit code is stable.

\subsection{2D results}

The 2D simulations of the benchmark problem are run paralleled in 4 or 8 nodes.
Square cells are used and z direction is uniformly divided into 64 cells. The
space and time steps are fixed to $\Delta x=0.02/64$m and $\Delta t_e=\Delta
t_i=0.5\times10^{-10}$s. The numerical schemes are chosen as follows: (1)
particles are moving in X-Y-Z coordinates; (2) weighting charge density and
interpolating the field in $z-r^2$ scheme; (3)
$\chi_{i+1/2,j}=max(\chi_{i,j},\chi_{i+1,j})$ and
$\chi_{i,j+1/2}=max(\chi_{i,j},\chi_{i,j+1})$; (4)the potential is logarithm
interpolated at the gap between the lower electrode and the outer cylinder; (5)
charge conservation scheme is used; and (6) voltage source is directly applied
to the electrode and no external circuit is considered. During most time of the
simulations, totally $3-8\times10^6$ super particles per species are traced.
The simulation will take 30 to 90 hours for one simulation in 4 nodes depending
on the specific numerical schemes.

\begin{figure}
\includegraphics[scale=0.5]{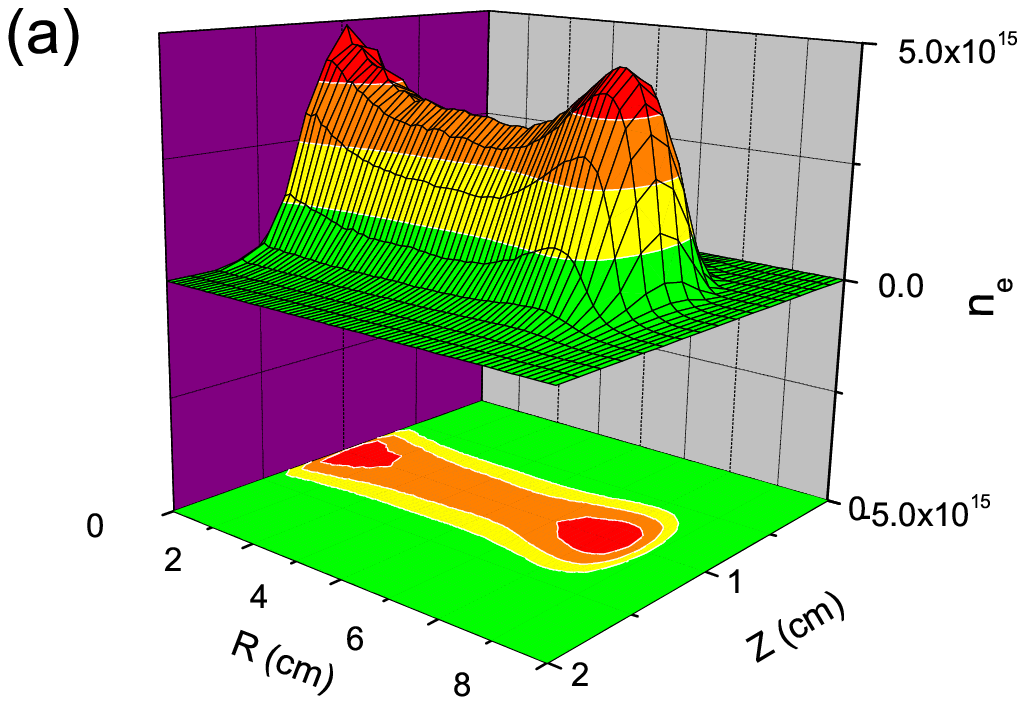}
\includegraphics[scale=0.5]{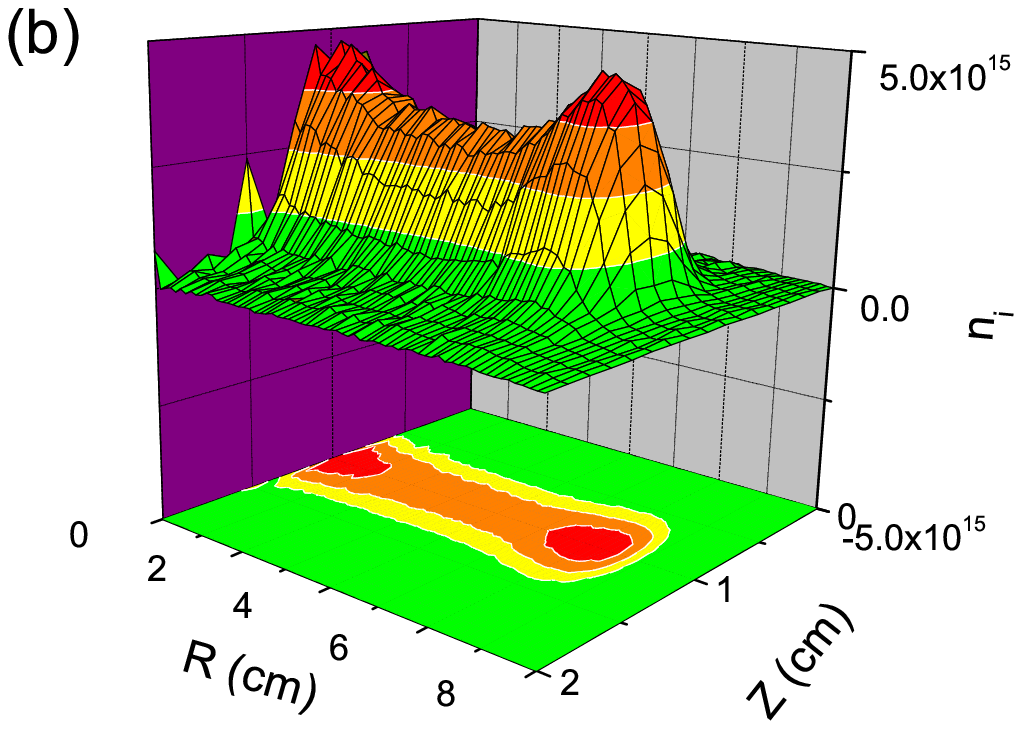}
\caption{\label{Fig2} 2D average electron(a) and ion(b) density
profiles.}
\end{figure}

The time averaging electron and ion density over one period are shown
Fig.\ref{Fig2}. The amplitude and profiles are consistent with the optical
emission tomography results \cite{Kitajima97}, the fluid simulations
\cite{Boeuf95} and the explicit simulation results \cite{Wakayama03}. The
density distribution along Z is very similar to the 1D results. There are two
densities peaks existing along R. One is near the axis and the other is near
the gap between the electrodes and the outer cylinder, formed a saddle-like
profile. This profile has also been observed in experiments \cite{Kitajima97}.
There are some noises in the axis of the ion density.

\begin{figure}
\includegraphics[scale=0.5]{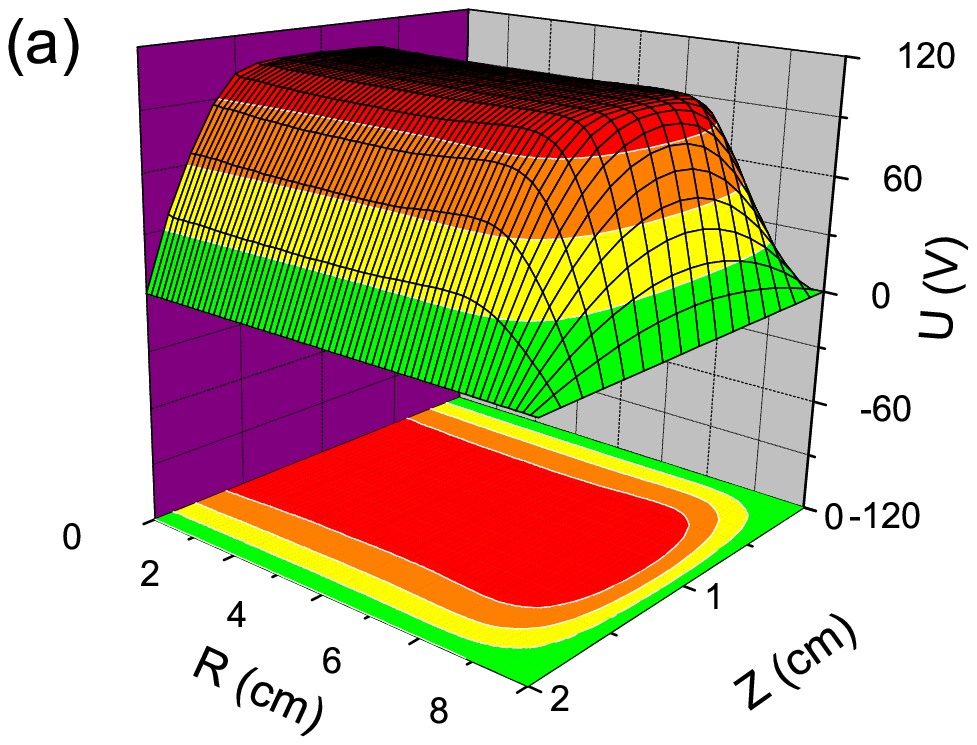}
\includegraphics[scale=0.5]{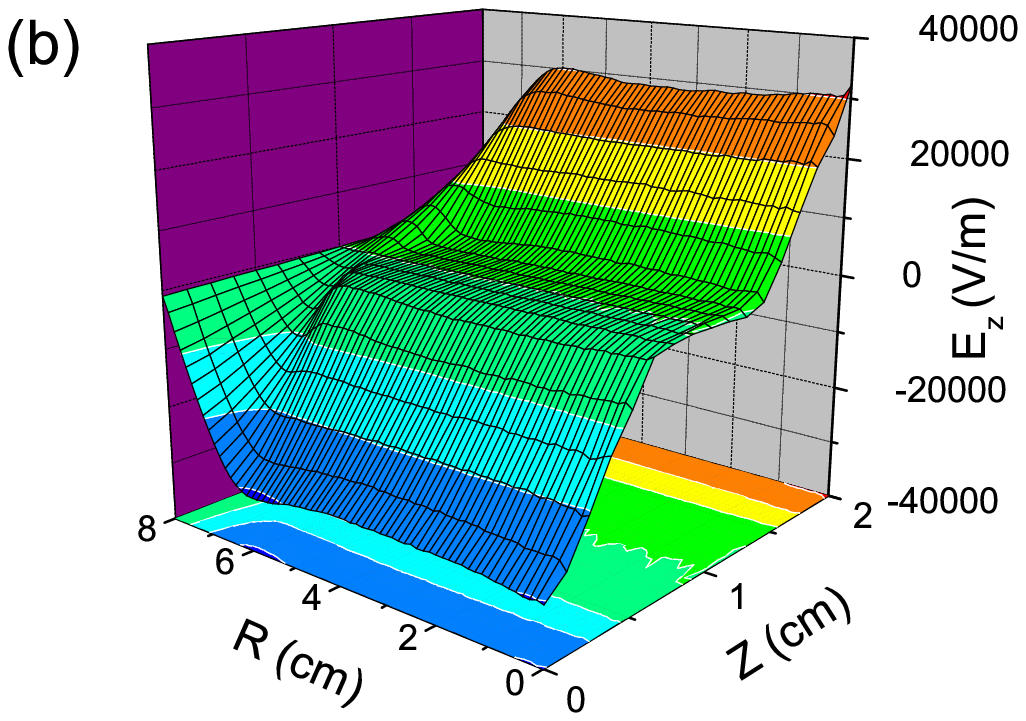}
\includegraphics[scale=0.5]{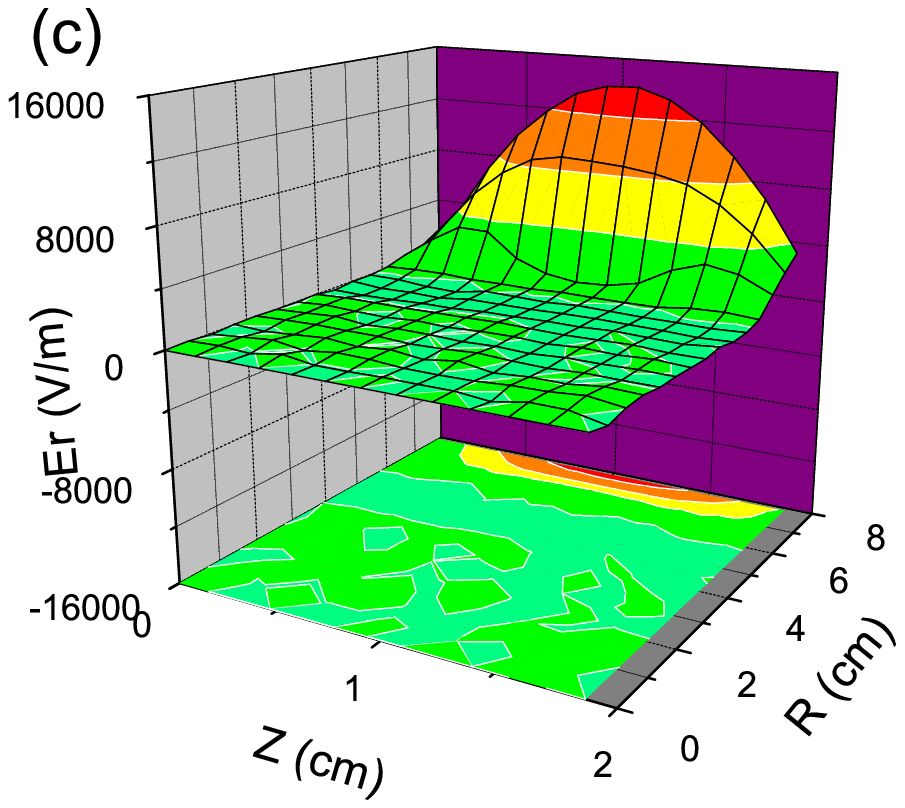}
\caption{\label{Fig3} 2D average (a)$\Phi$, (b)$E_z$ and (c)$E_r$
profiles.}
\end{figure}

The time averaging potential $\Phi$, $E_z$ and $E_r$ are illustrated in Fig.
\ref{Fig3}. It shows that $\Phi$ and $E_z$ have the profiles along Z similar to
the 1D results except for in the region near the gap. The $E_r$ is smaller than
$E_z$ and only obvious at large radius, because there is no rf voltage applied
and a radial sheath is formed.

\section{Discussion and summary}

In the present work, we have developed an implicit and electrostatic
Particle-in-cell/Monte Carlo model in two-dimensional axisymmetric geometry. We
discussed the available algorithms in the cylindrical implicit simulations in
detail. Benefits and shortcoming of several possible algorithms were analyzed
and compared to select the most reliable technologies. 1D and 2D benchmarks
were executed to validate the code and show the possible errors of the
algorithms. Although our code can be used to study most practical CCP devices
and the results seem satisfying, some issues still need to be addressed.

Although our MC code has included the model for O$_2$ and CF$_4$, simulation
for these electronegative gas is only possible for 1D case, for the electrons
and ions are equally weighted in our present model. One should adopt
species-depending weight scheme\cite{Nanbu00}, but we have not worked out all
the details for the MC process and the resizing method in this complex case.

Another problem should be mentioned is the smoothing. Filtering or smoothing
\cite{Birdsall85,Verboncoeur05} for the summed-up charge density can be used
before solving the Poisson equations. In Z direction, binomial filter can be
applied, and volume corrected filters can be used in R direction. However, we
find conventional filters have no significant effects, even up to 200 pass
being used for both $\rho$ and $\chi$. We believed the implicit method has
significantly damped the high frequency mode, so the smoothing should be
different from the implicit model.

The density in the axis shows $10\% \sim 30\%$ difference from the adjacent
line of grids. This phenomenon is also observed in DSMC simulations
\cite{Bird94, Liechty08} and is attributed to the numerical diffusion effects.
Furthermore, the density in the axis is very noisy. Although it seems that this
phenomenon has little effects on the final results. We are now trying to solve
this problem.

It seems the D1 implicit scheme is over damped, and thus produces smaller
density than explicit scheme in some cases. Our 1D benchmarks have shown that
adjustable damping schemes \cite{Friedman90} could give better results with
same space and time steps.

When incorporating external static magnetic field into this model, this model
can be used for many other similar devices, such as magnetrons and Hall
thrusters. The major differences are the particle moving and the field solver,
which have been studied before \cite{Brackbill85,Friedman90}. We are now trying
to improve the model by addressing above issues, and also adding more gas model
into our code.

\section*{Acknowledgments}
This work was supported by the National Natural Science Foundation of China
(No.10635010) and the Research Fund for Doctoral Program of Higher Education of
China (No. 20090041110026).

\end{document}